\documentclass[12pt]{article}
\usepackage{epsfig, amsmath, amssymb}
\usepackage{graphicx,psfrag} 
\newcommand{\be}{\begin{equation}}
\newcommand{\ee}{\end{equation}}
\newcommand{\ben}{\begin{equation}}
\newcommand{\een}{\end{equation}}
\newcommand{\bea}{\begin{eqnarray}}
\newcommand{\eea}{\end{eqnarray}}
\newcommand{\bA}{\begin{array}}
\newcommand{\eA}{\end{array}}
\newcommand{\bc}{\begin{center}}
\newcommand{\ec}{\end{center}}
\newcommand{\al}{\alpha}

\newcommand{\ra}{\rightarrow}

\newcommand{\ie}{{\it i.e.}}
\newcommand{\eg}{{\it e.g.}}

\newcommand{\ua}{\uparrow}
\newcommand{\da}{\downarrow}
\newcommand{\lan}{\langle}
\newcommand{\ran}{\rangle}

\begin{document}


\begin{titlepage}

%
  
\bc

\hfill 
\\         [30mm]

{\Huge de Sitter entropy as entanglement}
\vspace{16mm}

{\large K.~Narayan} \\
\vspace{3mm}
{\small \it Chennai Mathematical Institute, \\}
{\small \it H1 SIPCOT IT Park, Siruseri 603103, India.\\}

\ec
\vspace{40mm}

\begin{abstract}
We describe connected timelike codim-2 extremal surfaces stretching
between the future and the past boundaries in the static patch
coordinatization of de Sitter space.  These are analogous to rotated
versions of certain surfaces in the $AdS$ black hole. The existence of
these surfaces via the $dS/CFT$ framework suggests the speculation
that $dS_4$ is dual to two copies of ghost-like CFTs in a
thermofield-double-type entangled state. In studies of entanglement in
ghost systems and ``ghost-spin'' chains, we show that similar
entangled states in two copies of ghost-spin ensembles always have
positive norm and positive entanglement.
\end{abstract}

{\small \emph{Essay written for the Gravity Research Foundation 2019 Awards
    for Essays on Gravitation}  \vspace{-3mm}
\bc Submitted on 26 March 2019\ec }

\end{titlepage}



\section{On de Sitter entropy and $dS/CFT$}

de Sitter space is fascinatingly known to possess entropy
\cite{Gibbons:1977mu}\ (reviewed in \cite{Spradlin:2001pw}).
In the static patch coordinatization of $dS_{d+1}$\ (Figure-1), 
the Northern/Southern diamonds $N/S$ are static patches
with time translation symmetries. de Sitter entropy is
\be\label{dSent}
S_{dS_{d+1}} = {l^{d-1} V_{S^{d-1}}\over 4G_{d+1}}\qquad \xrightarrow{\ d\ra 3\ }
\qquad  {\pi l^2\over G_4} = S_{dS_4}\ ,
\ee
the area of the cosmological horizon (size $l$), apparently
stemming from degrees of freedom not accessible to observers in
regions $N/S$, for whom the horizons are event horizons.

The natural boundary for $dS$ is future/past timelike infinity $I^\pm$.
It is thus of interest to understand if this entropy can be realized in
gauge/gravity duality
\cite{Maldacena:1997re,Gubser:1998bc,Witten:1998qj,Aharony:1999ti} for
$dS$, or $dS/CFT$
\cite{Strominger:2001pn,Witten:2001kn,Maldacena:2002vr}, which 
conjectures $dS$ to be dual to a hypothetical Euclidean non-unitary
Conformal Field Theory that lives on ${\cal I}^+$,
with the dictionary $\Psi_{dS}=Z_{CFT}$\ \cite{Maldacena:2002vr}.\ \
$\Psi_{dS}$ is the late-time Hartle-Hawking Wavefunction of the
Universe with appropriate boundary conditions and $Z_{CFT}$ the dual
CFT partition function. The CFT$_d$ energy-momentum tensor $T_{ij}$
2-point correlators yield central charges ${\cal C}_d\sim
i^{1-d}{l^{d-1}\over G_{d+1}}$\,, effectively analytic continuations
from the anti de Sitter case: \eg\ for $dS_4$,
\be\label{dS4/CFT3}
Z_{{}_{CFT}}=\Psi_{_{dS}} \sim e^{iS_{cl}}\sim 
e^{-\int_k R_{dS}^2 k^3 \varphi_{-k}^0\varphi_k^0+\ldots}\ ,
\qquad\quad  \lan O_kO_{k'}\ran\sim
{\delta^2Z_{{}_{CFT}}\over\delta\varphi_k^0\delta\varphi_{k'}^0}\ ,
\ee
(semiclassically) for operators $O_k$ dual to modes $\varphi_k$.
The $\lan TT\ran$ correlator for $O_k$ being
appropriate $T_{ij}$ components gives the real, negative, central charge 
$-{R_{dS}^2\over G_4}$\,: thus
$dS_4/CFT_3$ is reminiscent of ghost-like ($c<0$) non-unitary
theories. In \cite{Anninos:2011ui}, a higher spin $dS_4$ duality was
conjectured involving a 3-dim CFT of anti-commuting (ghost) scalars,
which exemplifies this (see also \eg\
\cite{Strominger:2001gp}-\cite{Anninos:2017eib}).\
$dS/CFT$ duality, regardless of its existence, is perhaps useful to
organize our understanding of de Sitter space.
Bulk expectation values weighted by the bulk probability $|\Psi_{dS}|^2$
\cite{Maldacena:2002vr} are
\be
\lan\varphi_k\,\varphi_{k'}\ran \sim \int D\varphi\ \varphi_k\,\varphi_{k'}\,
\big|\Psi_{dS}[\varphi_k]\big|^2\ .
\ee
The existence of $\Psi$ and $\Psi^*$ here suggests that the dual
actually involves two copies of the CFT for a fixed $dS$ background\
(strictly one should also sum over all final 3-metrics $g^3$ in
$|\Psi_{dS}[\varphi,g^3]|^2$).

It is interesting to ask \cite{Narayan:2017xca} if de Sitter entropy
is some sort of generalized entanglement entropy via $dS/CFT$, viewed
from the future/past universes $F/P$ (Figure-1). From the bulk
perspective, a speculative generalization of the Ryu-Takayanagi
formulation \cite{Ryu:2006bv,Ryu:2006ef,HRT,Rangamani:2016dms}
involves the bulk analog of setting up entanglement entropy in the
dual theory.  In ordinary (static) quantum systems with spatial
subsystems on a constant time slice, entanglement entropy arises by
tracing over the environment. The dual theory here however is
Euclidean and spatial, with no intrinsic ``time''.  Operationally we
could define some spatial isometry direction as boundary Euclidean
time, then define a subsystem on this slice: this would lead to
codim-2 bulk extremal surfaces stretching in the time direction, if
they exist\ (all Euclidean time slices should be equivalent).  From
the boundary perspective, ghost-like CFTs as \cite{Maldacena:2002vr},
\cite{Anninos:2011ui}, might suggest, are expected to have negative
norm states/configurations, thus suggesting ``negative entanglement'':
it is interesting then to ask how a positive quantity like de Sitter
entropy might arise.

\section{Extremal surfaces}

In $AdS$, surfaces starting at the boundary dip into the radial
direction and exhibit turning points where they begin to return to the
boundary. In $dS$, the boundary at $I^+$ is spatial: surfaces dip into
the time direction giving a crucial minus sign that ensures that there
is no \emph{real} turning point where the surface starting at $I^+$
begins to turn back towards $I^+$.
There are also complex extremal surfaces with turning points, which
amount to analytic continuation from the $AdS$ Ryu-Takayanagi surfaces
\cite{Narayan:2015vda}-\cite{Miyaji:2015yva}.  While their interpretation
is not entirely clear, in $dS_4$ these have \emph{negative} area,
consistent with (\ref{dS4/CFT3}) for $dS_4/CFT_3$.

Since real surfaces starting at the future boundary $I^+$ sally forth
into the bulk without returning, it is interesting to ask if they
could instead end at the past boundary $I^-$\ \cite{Narayan:2017xca}.
The bulk probability $\Psi_{dS}^*\Psi_{dS}$ suggests two CFT copies:
so such connected extremal surfaces stretching between $I^\pm$ are
perhaps expected. Towards studying this, we recast $dS_{d+1}$ as
\vspace{-4mm}
\begin{figure}[h] 
\hspace{0.05pc}
\includegraphics[width=13pc]{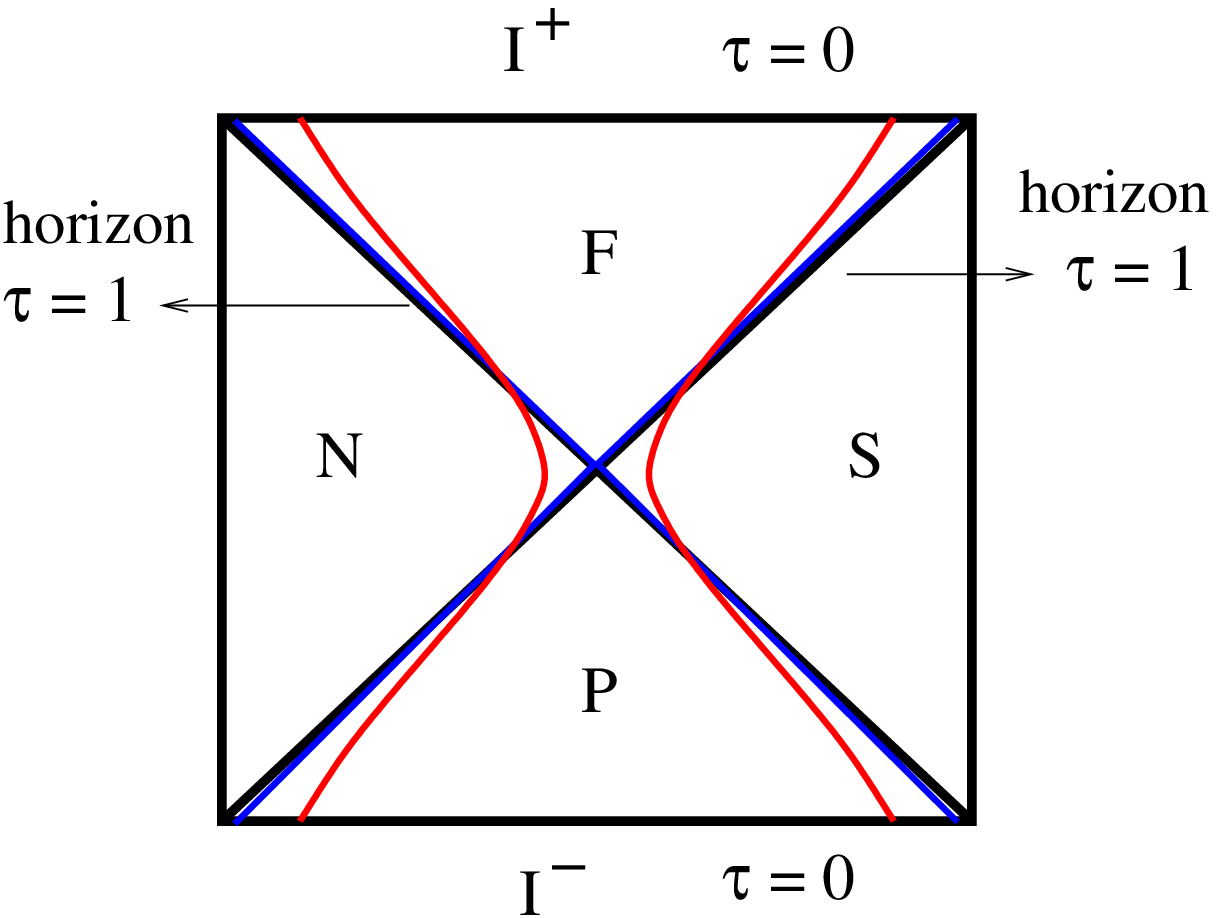} 
\hspace{2.3pc}
\begin{minipage}[b]{23pc}
\be\label{dSst}
ds^2 = {l^2\over\tau^2} \left(-{d\tau^2\over 1-\tau^2} + (1-\tau^2) dw^2
+ d\Omega_{d-1}^2\right) ,   
\ee
$\tau$ is ``bulk'' time in the future/past universes $F/P$\\
with $0\leq \tau\leq 1$. The horizons at $\tau=1$ have area
(\ref{dSent}).\\
$N/S$ have $1<\tau\leq\infty$ (with $w$ time).\ \ \ [Figure-1]\\
From the $\tau\sim 0$ asymptotics, the boundary is seen to\\
be Euclidean $R_w\times S^{d-1}$, while local patches are flat.
\end{minipage}
\end{figure}

The scaling ${l^{d-1}\over G_{d+1}}$ of de Sitter entropy
suggests codimension-2 surfaces. Likewise, entanglement in the dual theory
defined with respect to Euclidean time would suggest bulk surfaces
on appropriately defined constant boundary Euclidean time
slices of the bulk.  Given $w$-translation
symmetry and $S^{d-1}$ rotational invariance in (\ref{dSst}), we
restrict to a $w=const$ surface, or some $S^{d-1}$ equatorial plane 
(which are all equivalent).
For the latter, we expect extremal surfaces stretching from
a subsystem $\Delta w\times S^{d-2}$ at $I^+$ to an equivalent one at
$I^-$.\
Extremizing the area functional then gives the real surfaces $w(\tau)$,\
with\ $B^2=const$,
\be\label{HMsurf1}
         {\dot w}^2 \equiv (1-\tau^2)^2 \Big({dw\over d\tau}\Big)^2
         = {B^2\tau^{2d-2}\over 1-\tau^2
           + B^2\tau^{2d-2}}\ ,
\ee
${\dot w}$ in (\ref{HMsurf1}) is the
$y$-derivative, with $y=\int {d\tau\over 1-\tau^2}$ the ``tortoise''
coordinate, useful near the horizons.
For any finite $B^2>0$, we have ${\dot w}\ra 0$ near the boundary $\tau\ra 0$,
with ${\dot w}<1$ for $\tau<1$ (within $F$) and ${\dot w}\ra 1$ as
$\tau\ra 1$. The turning point $\tau_*$ is the ``deepest'' location
to which the surface dips into in the bulk, before turning around:
this is when
\be
 |{\dot w}|\ra\infty:\qquad\qquad
 1-\tau_*^2+B^2\tau_*^{2d-2}\ =\ 0\ .
\ee
Real $\tau_*(B^2)$ arises only if $\tau>1$ \ie\ within $N/S$. Overall
this gives the smooth ``hourglass''-like red curve in Figure-1 stretching
from $I^+$ to $I^-$, intersecting the horizons, turning around smoothly
at $\tau_*$ in $N$/$S$. These are akin to rotated versions of the
Hartman-Maldacena surfaces \cite{Hartman:2013qma} in the $AdS$ black
hole (which itself resembles a rotation of (\ref{dSst})).

The limit $B\ra 0$ gives $\tau_*\ra 1$, the turning point approaching
the bifurcation region (the red ``hourglass neck'' is pinching off).
The width becomes\ 
$\Delta w\sim \log{2\over\tau_*-1}\ra\infty$\ 
covering all $I^\pm$ (on this equatorial plane). In this limit
\cite{Narayan:2017xca} the area becomes   
\be\label{HMsurfArea}
S\ \ra\ {2 l^{d-1} V_{S^{d-2}}\over 4G_{d+1}}
\int_\epsilon^1 {d\tau\over\tau^{d-1}}\, {1\over \sqrt{1-\tau^2}}\quad
\xrightarrow{\ dS_4\ }\quad {\pi l^2\over G_4} {1\over\epsilon}\ ,
\ee
scaling as de Sitter entropy. In the $w=const$ slice, similar surfaces
(tricky in general) can be shown to exist with the same area when the
subregion covers all space.

\section{Entanglement in ghost systems}

In general, extremal surfaces appear tricky in de Sitter space, unlike
$AdS$: the surfaces here connecting $I^\pm$ are thus interesting.
These lie on some boundary Euclidean time slice (all of which are
equivalent). As the boundary subregion approaches all of $I^\pm$, they
pass through the vicinity of the bifurcation region. The area law
divergence scales as de Sitter entropy (\ref{HMsurfArea}).  The
existence of these surfaces suggests entanglement between the future
and past boundaries (see also \cite{Arias:2019pzy}). This cannot be
entanglement in the usual sense, the dual CFTs being Euclidean.
However boundary directions admitting translation symmetries can be
taken as Euclidean time, leading to generalizations of entanglement.

Motivated by (\ref{dS4/CFT3}) for $CFT_{F,P}$\ \cite{Maldacena:2002vr},
\cite{Anninos:2011ui}, entanglement in
ghost-like theories was studied \cite{Narayan:2016xwq} in toy 2-dim
ghost-CFTs using the replica formulation, giving $S<0$ for $c<0$
ghost-CFTs under certain conditions, and in (possibly more illuminating)
quantum mechanical toy models of ``ghost-spins'' via reduced density
matrices (RDMs).
We define a ghost-spin as a 2-state spin variable with indefinite
inner product $\gamma_{\al\beta}$ (akin to those in $bc$-ghost CFTs)
\be\label{gsNorm}
\lan\ua|\da\ran = 1 = \lan\da|\ua\ran\ ,\qquad
\lan\ua|\ua\ran = 0 = \lan\da|\da\ran\ ,
\ee
unlike $\lan\ua|\ua\ran = 1 = \lan\da|\da\ran$ for a single spin.
Redefined states $|\pm\ran$
satisfy $\gamma_{\pm\pm}=\lan\pm |\pm\ran=\pm 1$. A two ghost-spin state
then has norm
\be\label{norm2gs}
|\psi\ran=\psi^{\al\beta}|\al\ran|\beta\ran:\qquad
\lan\psi|\psi\ran=\gamma_{\alpha\kappa} \gamma_{\beta\lambda}
\psi^{\alpha\beta} {\psi^{\kappa\lambda}}^* = |\psi^{++}|^2+|\psi^{--}|^2
- |\psi^{+-}|^2 - |\psi^{-+}|^2\ .
\ee
Thus although states
$|-\ran$ have negative norm, the state $|-\ran|-\ran$ has positive
norm. For ghost-spin ensembles \cite{Jatkar:2016lzq}, generic
states exhibit novel non-positive entanglement stemming from negative
norm contributions. However ``correlated'' states
such as $\psi^{++}|+\ran|+\ran+\psi^{--}|-\ran|-\ran$ entangling identical
ghost-spins between two copies of ghost-spin ensembles can be shown
to have positive norm, RDMs and entanglement. In
\cite{Jatkar:2017jwz}, 1-dim ghost-spin chains with specific
nearest-neighbour interactions were found to yield $bc$-ghost CFTs in
continuum limits, suggesting that ghost-spins are perhaps microscopic
building blocks of ghost-like non-unitary CFTs. Various $N$-level
generalizations of these 2-level ghost-spins \cite{Jatkar:2018lmt}
also exhibit similar correlated states. Thinking thereby of appropriate
3-dim $N$-level ghost-spin systems as microscopic realizations in the
universality class of ghost-$CFT_3$'s dual to $dS_4$ with
$N \sim {l^2\over G_4}$\ finite albeit large, consider
two ghost-CFT copies and a ``correlated'' state \cite{Narayan:2017xca}
\be\label{dsEntstate}
|\psi\rangle = \sum \psi^{i_n^F,i_n^P} |i_n\ran_{{}_F} |i_n\ran_{{}_P}\ ,
\ee
entangling generic ghost-spin
configurations $|i_n\ran_{{}_F}$ from $CFT_F$ at $I^+$ with
\emph{identical} ones $|i_n\ran_{{}_P}$ from $CFT_P$ at $I^-$, as the
$I^\pm$-connecting surfaces suggest. These are entirely positive, the
doubling cancelling the minus signs, giving positive RDM and entanglement.
In toy examples of two $N$-level ghost-spin chain copies
\cite{Jatkar:2018lmt}, the $N$ internal possibilities restricting to
ground states gives (maximal) entanglement entropy scaling as\ $N\sim
{l^2\over G_4}$\,.

Bulk time evolution, mapping states at $I^-$ to $I^+$\
\cite{Witten:2001kn}, suggests the states (\ref{dsEntstate}) are
unitarily equivalent to entangled states
in two $CFT_F$ copies solely at $I^+$.  While $CFT_P\equiv CFT_F$
implies a single CFT, the state (\ref{dsEntstate}) is best regarded as
a particular entangled slice in a doubled system.  Thus
(\ref{dsEntstate}) is akin to the thermofield double dual to the
eternal $AdS$ black hole \cite{Maldacena:2001kr}. This suggests the
speculation that $dS_4$ is perhaps approximately dual to $CFT_F\times
CFT_P$ (or $CFT_F\times CFT_F$) entangled as (\ref{dsEntstate}),
$dS_4$ entropy arising as entanglement. Many issues arise of course
(see the overview \cite{StromingerStrings2012}):
\eg\ on regions $N, S$ (which are ``behind the horizons'' viewed from
$dS/CFT$ at $I^\pm$) and interior operators, akin to \eg\
\cite{Papadodimas:2013jku} for black holes.

\vspace{7mm}

{\footnotesize \noindent {\bf Acknowledgements:}\ \ It is a pleasure
  to thank Dileep Jatkar for discussions and collaboration on some of
  this work. I also thank Sumit Das, Shiraz Minwalla and Sandip
  Trivedi for useful recent discussions. This work is partially
  supported by a grant to CMI from the Infosys Foundation.  }


\end{document}